\documentclass[12pt,preprint]{aastex}
\usepackage{natbib}
\begin{document}

\title{Exponential Galaxy Disks from Stellar Scattering}

\author{Bruce G. Elmegreen}
\affil{IBM Research Division, T.J. Watson Research Center, 1101 Kitchawan Road, Yorktown
Heights, NY 10598, bge@watson.ibm.com}

\and

\author{Curtis Struck}

\affil{Department of Physics and Astronomy, Iowa State University, Ames, IA 50011,
curt@iastate.edu}

\begin{abstract}
Stellar scattering off of orbiting or transient clumps is shown to lead to the
formation of exponential profiles in both surface density and velocity dispersion
in a two-dimensional non-self gravitating stellar disk with a fixed halo
potential. The exponential forms for both nearly-flat rotation curves and
near-solid body rotation curves.  The exponential does not depend on initial
conditions, spiral arms, bars, viscosity, star formation, or strong shear. After a
rapid initial development, the exponential saturates to an approximately fixed
scale length. The inner exponential in a two-component profile has a break radius
comparable to the initial disk radius; the outer exponential is primarily
scattered stars.
\end{abstract}


\section{Introduction}
Spiral galaxies have exponential radial profiles \citep{freeman70, vanderkruit02}
with a nearly uniform central surface brightness for early and intermediate Hubble
types and a trend toward lower central brightnesses for later types
\citep{jong96,graham01}. The scale length is also independent of Hubble type for
early and intermediate types \citep{jong96}, but increases with increasing galaxy
mass \citep{fathi10} within those types. Generally galaxies with larger scale
lengths have fainter central brightnesses \citep{fathi10}.

The radial profiles themselves show distinct types, originally classified by
\cite{freeman70} and studied more recently by \cite{pohlen06}, \cite{erwin08},
\cite{herrmann13} and others.  Type I is a uniform exponential with a constant
scale length, Type II is a double exponential with a steeper outer part, and Type
III is a double exponential with a shallower outer part. These types correlate
with Hubble type in the sense that Type IIs are more common in late type galaxies
\citep{pohlen06, gutierrez11, herrmann13}. Late types also have a smaller ratio of
the break radius to the inner scale length \citep{pohlen06, he06}.

The are several theories for the origin of the exponential shape. A uniform-density,
uniformly rotating spherical halo that collapses to a disk with no angular momentum
redistribution between mass elements \citep{mestel63} has a distribution of mass as a
function of angular momentum that is nearly the same as that of a self-gravitating
exponential profile out to a radius $R\sim6R_{\rm D}$ scale lengths \citep{freeman70}.
Early observations of outer disk angular momentum distributions supported such idealized
collapse \citep{crampin64} as did early models of disk formation \citep{fall80}.
\cite{efstathiou00} showed, however, that no exponential formed this way can go out very
far if it comes from a power law halo because the outer disk mass is exponentially small
and has to come from a very narrow range of halo radii where the range of angular momenta
is too narrow compared to that required for the disk \citep{ferguson01}. Now we know that
some galaxies have exponential disks out to $\sim10R_{\rm D}$
\citep[e.g.,][]{weiner01,bland05}.

The \cite{mestel63} model was popular when exponential disks seemed to have abrupt
outer truncations at $\sim4.5R_{\rm D}$ \citep[e.g.,][]{dalcanton97}, in which
case this outer radius could be related to the maximum angular momentum of the
initial halo gas. Observations now show a transition to a second exponential or
other peripheral stellar structure at about this radius with no sharp truncation
even in extremely deep surveys based on photometry \citep[e.g.,][]{hunter11} or
star counts \citep{saha10,mcconnachie10,vlajic11,grossi11,radburn12, barker12}.
The assumption of specific angular momentum conservation is also not made anymore
because tidal torques, spiral arms, bars, mergers and continuing gas accretion
redistribute the disk angular momentum.

A second model proposes that the exponential shape forms through secular processes
when the viscous accretion rate in the disk is proportional to the star formation
rate \citep{lin87, yoshii89,ferguson01,wang09}. This model requires shear and
works for spiral galaxies, but not for dwarf Irregulars that have exponential or
double-exponential disks like spirals, and sometimes out to $\sim 8 R_{\rm D}$ as
well \citep{hunter11}, but nearly solid body rotation in at least the inner parts.
\cite{he06} also pointed out that there is no obvious relation between the break
radius in a dwarf Irregular and the turn-over radius of the rotation curve.

A third model involves a galaxy bar, which redistributes disk angular momentum and
can make an exponential from some initially different distribution \citep{hohl71}.
Bars also make double exponentials with the outer part just beyond the outer
Lindblad resonance of the spirals that the bar generates
\citep{debattista06,foyle08}. Bars are not present in all disks, however, nor are
there bulge-like remnants of bars in late type galaxies even though late types
dominate those with double exponentials \citep{pohlen06}. Neither is there a
correlation between the presence of bars and exponentials or double exponentials
in dwarf Irregular galaxies \citep{he06}. High redshift galaxies, out to $z=5.8$
in \cite{fathi12}, have exponential disks as well, so this shape has to arise
quickly.

Additional models consider possible origins for the break in a single exponential,
making it double, such as a threshold in star formation at the break radius
\citep{fall80, foyle08}, which either terminates the disk at that point or makes
an outer exponential that falls more steeply than the inner \citep{eh06}. Grazing
encounters with other galaxies can make a double exponential with the outer one
more shallow \citep{younger07, penarrubia06}. Models with firm star formation
thresholds require a mechanism to get stars beyond the threshold, and a good
candidate for such a mechanism is star scattering by spiral arms
\citep{sellwood02, roskar08, martinez09}. Spiral scattering with a star formation
threshold makes the outer disk come from the inner disk, and it predicts a color
profile that gets bluer with increasing radius in the star formation part and then
redder with radius outside of that in the scattered part. Such color profiles are
observed in both local \citep{bakos08} and redshift $\sim1$ galaxies
\citep{azz08}. However, two-component surface brightness profiles with such a
color distribution convert into single-exponential mass profiles that seem too
regular to be explained by a two-component (star formation plus scattering) model
\citep{bakos08}. Also dwarf Irregular galaxies with double exponentials have no
spiral arms to do the scattering. There are problems with the encounter theory as
well: \cite{gutierrez11} point out that Type III's are not just Type I's with
their outer parts spread out because the III's are brighter inside.

Numerical simulations of galaxy formation get exponential and double-exponential
disks under realistic cosmological conditions, so we might ask how they do it.  An
essential ingredient for a realistic disk is star formation feedback
\citep{robertson04}, but the primary role of this feedback is to prevent the gas
disk from getting too small. The actual exponential shape that results in these
and other simulations is not specifically questioned. Simulations with double
exponentials produce them from initial single exponentials, but again the
exponential shape itself is not explained. Perhaps the exponential results from a
combination of processes listed above, even though no single mechanism can be
applied to all galaxies. Or, perhaps there is an additional ingredient in these
simulations and in real galaxies that has not been specifically noted.

The most stringent test for exponential models seems to be dwarf Irregular
galaxies, as noted above. They have little shear, no spirals, are often
non-barred, and yet have an underlying disk plus star formation that follows an
exponential distribution out to 6 or more scale lengths \citep[to 30
mag/arcsec$^2$ in V-band;][]{hunter11} -- long into the regime where the disk
seems to be gravitationally stable \citep{vanzee97, hunter11}. They are also
harassed by encounters with galaxies in their groups or by smaller companions
\citep{pustilnik01}, and many could be subject to continuing gas accretion
\citep{vanzee98,wilcots98}.  Star formation sometimes mimics the underlying
exponential so closely \citep[also in spirals;][]{shi11, hunter13} that it is
unclear whether the total stellar exponential is made from the accumulated history
of star formation at each radius, or if star formation follows the stars which
take an exponential form for other reasons. As discussed in \cite{hunter13}, the
first case seems unlikely because then the dense molecular gas that forms stars
has to know about the gas distribution in the whole disk; the second case is more
reasonable as it requires the stars to continuously migrate to smooth out
non-exponential perturbations. The problem is that in dwarf Irregulars, the stars
have to do this migration without spirals, bars or significant shear.

Here we run simple experiments of passive stellar scattering in rotating thin
disks that are subject to steady or transient perturbations from point-like
gravitating objects.  Such objects could be giant star-forming regions in the case
of dwarf Irregulars \citep{hunter11} or they could be clumps in high-redshift
galaxies \citep{bournaud07}. They could also be dense halo objects like
dark-matter minihalos that pass through the disk. We find that perturbations like
this can make an exponential surface density distribution from an initially flat
distribution, and that the distribution of stellar velocity dispersion also
becomes exponential from an initially uniform value, with a scale length somewhat
longer than for the mass. This result for the dispersion is interesting because it
makes the scale height more constant than it would be with the surface density
decreasing alone. A near-uniform disk scale height is another observational
constraint on the origin of exponential disks \citep[e.g.,][]{degrijs97}.

\section{Models}

The simplest model for disk evolution with stellar scattering consists of
two-dimensional orbital integrations of non-interacting test particles (stars) in
centrifugal balance in a fixed potential (halo), along with softened, point-mass
gravitating particles (clumps) that are also in the disk. Time integrations are
carried out with the MATLAB Runge-Kutta routine "ode23." The lack of self-gravity
in the stars means that resolution effects are not important. We consider two
cases, one with a nearly flat rotation curve and another with a nearly solid-body
rotation curve.

\subsection{Nearly-flat rotation curve}

For the first case, the gravitational acceleration of the fixed halo is of the form
\begin{equation}
g_{\rm F}(R)=-{{GM_{\rm H}}\over{H^2}}\left({{R}\over{H}}\right)^{-1.1}
\end{equation}
where $H$ is the halo scale length, and $M_{\rm H}$ is the halo mass within $H$.
This acceleration gives a rotation curve that is slightly falling, $V(R)\propto
R^{-0.05}$. The clumps in this model have a softening length $0.01H$ and are on
fixed circular orbits without responding to the stars. The initial disk surface
density is independent of radius and has a sharp outer edge at $R=4$ units, as do
the clumps. The sharp edge in the stellar distribution does not affect the
dynamics because the stars are not self-gravitating.

We take dimensionless length and time units, $H=1$ and $T=H/V_{\rm circ}=1$ where
$V_{\rm H}^2=GM_{\rm H}/H$. As an example of a large spiral galaxy like the Milky
Way, we consider $H=2$ kpc and $V_{\rm H}=220$ km s$^{-1}$. Then $T=8.9$ Myr,
$M_{\rm H}=HV_{\rm H}^2/G=2.2\times10^{10}\;M_{\odot}$, and the rotation period at
$R=H$ is $2\pi$ code units or $56$ Myr. At $R=10$ kpc, the rotation period is
$303$ Myr.

The left-hand side of Figure 1 shows the results of a simulation with 15925
non-interacting stars in initially circular orbits with zero velocity dispersion
and with 70 gravitating clumps of mass $0.002M_{\rm H}$ each in fixed circular
orbits. The figure shows the distribution of these stars (top panel), the radial
dependence of stellar surface density (middle panel), and the radial dependence of
the rms stellar velocity dispersion (bottom panel) after 300 time units
(corresponding to 2.7 Gyr). The rms velocity dispersion is measured as
$\sigma_{\rm rms}=(\sigma_{\rm r}^2+\sigma_{\theta}^2)^{1/2}/2^{1/2}$ for radial
and angular components $\sigma_{\rm r}$ and $\sigma_{\theta}$; it is normalized to
$V_{\rm H}$.

The disk has evolved to have exponential profiles in surface density and velocity
dispersion. Without perturber clumps (Fig. 2, black curve), there is no
significant evolution of the disk, which keeps its initially flat surface density
profile out to the sharp edge. Plots (not shown) of individual stellar
trajectories indicate that the co-rotating clumps are scattering stars as they
shear by. This scattering drives the overall evolution of the disk toward an
exponential profile.  The profile is actually double-exponential with the inner
part comparable in size to the initial disk.

The exponential decline of the velocity dispersion in Figure 1 suggests an
explanation for the nearly constant scale height of stellar disks as a function of
radius \citep[e.g.,][]{degrijs97}. The scale height depends on the ratio of the
square of the perpendicular velocity dispersion to the surface density. If the
radial dispersion induced by clump scattering in our model converts in part to a
vertical dispersion through initial and subsequent scattering, and the vertical
dispersion also takes on an exponential profile of the same type, then the
vertical scale height would be more constant than in the case of an isothermal
disk.

The formation of the exponential profile occurs steadily over time. The evolution
is faster if the total mass in perturbing clumps is larger. Eventually the profile
saturates, evolving very slowly thereafter. This evolution is shown in Figure 2,
which plots the surface density and rms velocity dispersion versus radius for 4
different times. The inner exponential has been established by 100 time units and
hardly evolves after that. The saturation occurs when every stellar scattering
outward is approximately matched by a stellar scattering inward.  An example of
saturation may occur in the self-gravitating N-body simulation by \cite{donghia13}
where a single massive perturber present for a short time in an already
established exponential disk did not change the radial profile noticeably.

\subsection{Near-solid body rotation}

In the case of near-solid body rotation, the gravitational acceleration of the
fixed halo is taken to be of the form
\begin{equation}
g_{\rm SB}(R)=-{{GM_{\rm H}}\over{H^2}}\left({{R}\over{H}}\right)^{0.8}.
\end{equation}
The dimensionless length and time are defined in the same way as for the
nearly-flat rotation case. Scaling to a dwarf irregular galaxy as an example with
this type of rotation curve, we take $H=0.5$ kpc and $V_{\rm H}=50$ km s$^{-1}$,
so that $T=9.8$ Myr, $M_{\rm H}=2.9\times10^{8}\;M_{\odot}$, and the rotation
period at $R=H$ is $61$ Myr.

The right-hand side of Figure 1 shows a simulation with 20187 non-interacting
stars in initially circular orbits with zero initial velocity dispersion and no
density gradient. The initial disk size is $9$ length units. There are 6
perturbing clumps also inside a radius of 9 units that are in fixed circular
orbits with a fractional mass of $0.15M_{\rm H}$ each. We pick fewer and more
massive clumps in the near-solid body case because dwarf galaxies are smaller
compared to the size of a typical star complex than are large spiral galaxies. The
softening parameter for the clumps is $0.01H$. Recall that the total mass of the
galaxy is larger than the scale mass $M_{\rm H}$, which is only that part inside
the halo scale length $H$.

To prevent a single clump from dominating a large number of stellar orbits, we
make each clump disappear after 10 time units (98 Myr) and reappear instantly at
some random azimuthal position with the same radius.  This hopping time is about
the timescale for giant star formation complexes to form and disperse. Random
clump appearance is also consistent with a model in which dark minihalos cross the
disk at random places and times.

The results shown in the figure correspond to the state of the disk after 50 time
units (490 Myr for the dwarf galaxy scaling). As for the case with a flat rotation
curve, the disk with a nearly solid body rotation curve also evolves toward an
exponential profile in surface density. The velocity dispersion profile is
exponential too, although slightly more flat than in the flat rotation curve case.

Both simulations shown in Figure 1 produce exponential disks, but a near-solid
body case tends to form a weak bar and the nearly-flat rotation curve case does
not. A bar is a quasi-static configuration for aligned elliptical orbits with a
wide range of radii in the solid body case. The perturbers tend to kick stars into
such elongated orbits and if enough stars respond to the same perturber, then
these stars combine to make an elongation that looks like a bar. While this effect
might be reasonable for dwarf irregular galaxies, we chose to mitigate it here by
randomizing the perturbations in azimuth. We make this choice because we already
know that bars can drive a disk toward an exponential profile, and we want to
isolate the effects of clump-like perturbations instead. The bars in our
simulations would not promote an exponential disk anyway because the stars that
make up the bars are not gravitating.

\section{Discussion}

All of our simulations with sufficiently strong clump-like perturbations drive an
initially uniform and cold stellar disk to an exponential form in both surface
density and velocity dispersion. The timescale for this evolution depends on the
strength of the clump forcing, but for a total relative clump mass of several tens
of percent, the time for significant evolution is less than a Gyr. Test cases with
no clumps confirmed that the initial profile was stable.

Models with a single massive clump (not shown) required a clump mass equal to
$\sim1$\% of the halo mass inside 10 radial units to modify the stellar disk in a
reasonable time. Such a massive clump also imprints a permanent asymmetry on the
disk, forcing the stars into a banana-shaped region centered on the potential
center, with a gap around the clump. Like the classical restricted three-body
problem, many of those stars follow banana-orbits in the co-rotating frame. Stars
originally near this clump scattered to larger radii on high eccentricity orbits
and eventually formed an extended disk with a roughly exponential profile over a
narrow range of radii (a few length units). The other stars in the disk retained
their initially flat profile because they were too distant from the clump to
experience significant scattering.  This result suggests that each clump
contributes independently to the exponential disk, and the ensemble of clumps is
what makes the final exponential symmetric.

Randomly appearing and disappearing clumps can stir a stellar disk more than
co-rotating clumps because the time-changing potential from the clumps acts like a
non-adiabatic heat source for all of the neighboring stars. This is presumably why
the near-solid body model makes an exponential quickly even though the local rate
of shear is low. The implication is that star formation clumps or transiting dark
mini-halos in dwarf Irregular galaxies promote the evolution of their stellar
disks into an exponential form. The same may be true in spiral galaxies, but
transient spiral arms may dominate the forcing there.

\section{Conclusions}

Stellar scattering off orbiting or transient clumps can produce exponential
profiles in two dimensional stellar disks with a fixed halo potential. No other
ingredients are necessary to produce the exponential, making this an attractive
solution to the problem of its origin, especially for dwarf Irregulars where there
are no spiral arms. Other processes that drive disk evolution, not modeled here,
should operate at the same time, but the ubiquity of exponentials from scattering
alone ensures that an exponential will eventually develop as long as these other
processes do not disrupt it.

We are grateful to D. A. Hunter for comments on the manuscript.

\clearpage
%fig1
\begin{figure}
\centering
\includegraphics[width=3.8in]{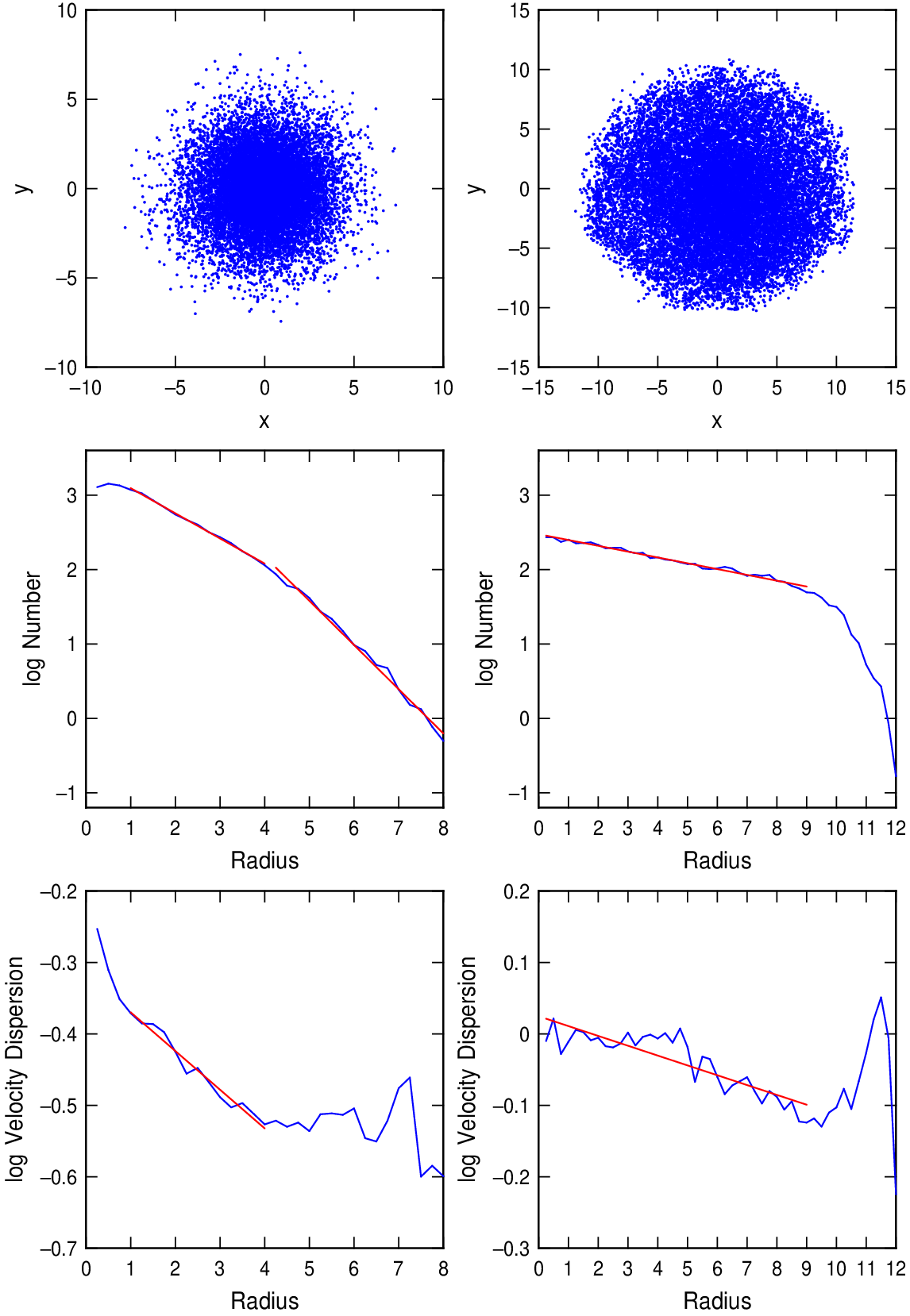}
\caption{(left) The stellar distribution (top), surface density profile (middle)
and velocity dispersion profile (bottom) for a non-self gravitating, two-dimensional model stellar disk
with a nearly-flat rotation curve imposed by a fixed potential.
The surface density is initially flat out to $R=4$ length units (8 kpc
if scaled to a Milky Way-size galaxy) and
then evolves to an exponential by the time shown, which is 300 time units ($\sim2.7$ Gyr).
The velocity dispersion starts at zero and also evolves to an exponential.
This evolution is driven by 70 perturbing clumps in circular orbits with a relative mass of 0.2\% each.
The fitted exponentials (red lines) have scale lengths of 3.0 and 1.7 length units
for the inner and outer parts of the surface density, and 18.5 length units for
the inner part of the velocity dispersion. (right) The same quantities as on the left
but now for a near-solid body rotation curve.
The initially flat and cold disk extends out to $R=9$ length units (4.5 kpc
for typical scaling of a dwarf Irregular galaxy). The
distributions shown are for 50 time units ($\sim0.49$ Gyr).
In this case, the evolution is driven by 6 perturbing clumps in circular orbits with a
relative mass of 15\% each; the clumps jump randomly in azimuth every 10 time units (98 Myr)
to minimize the formation of a bar.
The fitted exponentials have scale lengths of 12.8 units for surface
density and 73 units for velocity dispersion. } \label{exponential_figs}
\end{figure}

\clearpage
%fig2
\begin{figure}
\centering
\includegraphics[width=4.in]{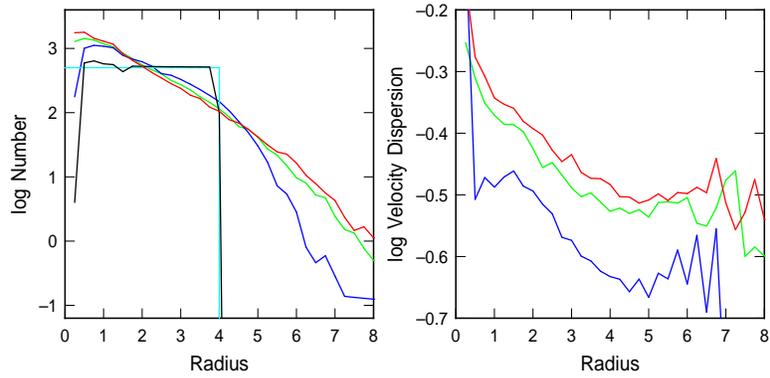}
\caption{(left) Surface density profiles for the nearly-flat rotation curve case
at times of 0 (cyan curve), 100 (blue), 300 (green), and 500 (red)
time units (0, 890, 2700, and 4450 Myr for the
Milky Way scaling). The time 300 case is the same as in figure 1.  The black curve
is for the same initial conditions but without perturbers, shown after 100 time
units. (right) The profiles of rms velocity dispersion at 100,
300, and 500 time units.} \label{exponential_fig2}
\end{figure}

\end{document}